\documentclass[aps,prl,showpacs,twocolumn,superscriptaddress]{revtex4}
\usepackage{graphics}
\usepackage{graphicx}
\usepackage{amsmath}
\usepackage{epsfig}

\usepackage{comment}
\usepackage{float}

\newcommand{\vect}[1]{{\bf {#1}}}

\begin{document}

\title{Superexchange-Driven Magnetoelectricity in Magnetic Vortices}

\author{Kris~T.~Delaney}
\affiliation{Materials Research Laboratory, University of California, Santa Barbara,
California 93106-5121, USA}
\author{Maxim~Mostovoy}
\affiliation{Zernike Institute for Advanced Materials, University of Groningen, The Netherlands}
\author{Nicola~A.~Spaldin}
\affiliation{Materials Department, University of California, Santa Barbara,
California 93106-5050, USA}

\date{\today}

\begin{abstract}
We demonstrate that magnetic vortices in which spins are coupled to polar lattice
distortions via superexchange exhibit an unusually large linear magnetoelectric
response. We show that the periodic arrays of vortices formed by frustrated spins
on Kagom\'e lattices provide a realization of this concept; our {\it ab initio}
calculations for such a model structure yield a magnetoelectric coefficient that is 30 
times larger than that of prototypical single phase magnetoelectrics. Finally, we 
identify the design rules required to obtain such a response in a practical material.

\end{abstract}


\maketitle


The ability to control magnetism with electric fields, which can
be realized through the interplay between spins and charges in solids, has an 
obvious technological appeal. The simplest form of such a control is the linear
magnetoelectric effect, when an antiferromagnet placed in an electric field $\vect{E}$
becomes a magnetized with magnetization $\vect{M}$, while an applied magnetic
field $\vect{H}$ induces an electric polarization $\vect{P}$, proportional to the field:
\begin{eqnarray}
\left\{
\begin{array}{rcl}
P_{i}&=&\alpha_{ij}H_j \\
M_{j}&=&\alpha_{ij}E_i \quad.
\end{array}
\right.
\end{eqnarray}
Here $\alpha_{ij}$ is the magnetoelectric tensor and summation over repeated indices is implied.
This magnetoelectric response requires simultaneous breaking of inversion and 
time-reversal symmetries, which defines the allowed magnetic symmetry classes and the non-zero components
of the magnetoelectric tensor.  While the phenomenology 
of the linear magnetoelectric effect is now well 
understood\cite{Landau/Lifshitz:Book,Fiebig:2005}, the use of magnetoelectrics is hampered 
by rather low values of their magnetoelectric constants: for example in the prototypical
magnetoelectric Cr$_2$O$_3$, a (large) electric field of $1 \times 10^6$ V/cm induces a (tiny) 
magnetization of $\sim 9 \times 10^{-5} \mu_B$ per Cr ion.
The search for materials with a much stronger response requires a deeper understanding of the
microscopic mechanisms of magnetoelectric coupling and new ideas about spin orders and crystal lattices
that can conspire to produce a large magnetoelectric effect.

The aforementioned technological driver is likely also responsible for the renewal of interest in the related class of
{\it multiferroic} materials which have simultaneous ferroelectric and (ferro)magnetic orders. Two recent 
developments in this field are particularly relevant for the work we will present in this Letter. First, 
an early observation that spiral magnetic order can lead to an electrical 
polarization\cite{Newnham/Kramer/Schulze/Cross_1978}, has been confirmed 
repeatedly\cite{Kimura_et_al:2003,Goto_et_al:2004} and the list of such materials has been considerably 
enlarged.
While spectacular non-linear magnetoelectric effects, such as reorientation of electric polarization
with a magnetic field, have been observed, the polarizations in such spiral magnets are small because 
the spin-lattice interaction is the weak spin-orbit-driven Dzyaloshinskii-Moriya 
interaction\cite{Katsura/Nagaosa/Balatsky:2005,Sergienko/Dagotto:2006}.  
At the same time, a new class of multiferroics has been identified in which the magnetic ordering couples 
to the lattice through mechanisms of non-relativistic origin, in particular exchange striction arising 
from superexchange\cite{Sergienko/Sen/Dagotto:2006,Picozzi_et_al:2007}. The stronger spin-lattice
coupling leads to correspondingly larger magnetically induced ferroelectric polarizations, with
polarization values close to those of conventional ferroelectrics suggested.

In this Letter, we show that these two concepts from the field of multiferroics -- symmetry breaking in 
spiral magnets, and super-exchange mediated spin-lattice coupling -- can be combined to yield materials 
with strong linear magnetoelectric response.

\begin{figure}[htbp]
\begin{center}
\resizebox{1.0\columnwidth}{!}{\includegraphics{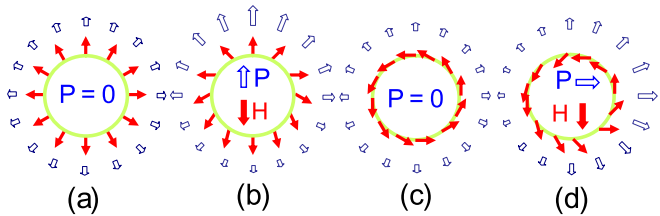}}
\end{center}
\caption{(a) A magnetic vortex carrying a pseudoscalar moment. The thin solid arrows indicate the spin orientation, 
while the thick open arrows show the local polarization vector; (b) A magnetic field applied to the vortex shown in 
(a) induces a net polarization along the field direction; (c) A magnetic vortex carrying a toroidal 
moment; (d) A magnetic field applied to the vortex shown in (c) induces an electric polarization perpendicular
to the field.} 
\label{fig:vortex}
\end{figure}

We begin by considering the magnetoelectric response of a single spin vortex (Fig.~\ref{fig:vortex}a). 
This can be viewed as a magnetic spiral rolled into a circle, and so we can use the results from spiral
multiferroics to analyze its magnetoelectric response. The magnetically induced ferroelectric polarization 
in spiral multiferroics is described by 
\begin{equation}
\vect{P} \propto \vect{e} \times \vect{Q} \quad, 
\end{equation}
where $\vect{e}$ is the axis around which the spins rotate and $\vect{Q}$ is the spiral
wavevector\cite{Mostovoy:2006}.
In our context, this coupling induces an inhomogeneous electric polarization locally oriented along the radial 
direction, so that the net polarization of the vortex is zero. A magnetic field applied in the $xy$ plane leads 
to a non-uniform rotation of spins in the vortex, which results in a nonzero net electric polarization proportional 
to the magnetic field (see Fig.~\ref{fig:vortex}b). 
The spin vortex shown in Fig.~\ref{fig:vortex}a has a diagonal magnetoelectric tensor, with magnetization 
induced parallel to the applied electric field, while for the vortex shown in 
Fig.~\ref{fig:vortex}c an applied magnetic field induces a perpendicular electric
polarization and the magnetoelectric tensor is antisymmetric (see Fig.~\ref{fig:vortex}d).
These conclusions are actually independent of the mechanism of magnetoelectric coupling
and generally follow from the fact that the vortices shown in Figs.~\ref{fig:vortex}a and
\ref{fig:vortex}c have respectively a monopole moment
$A \propto \sum_{\alpha} {\bf r}_{\alpha} \cdot {\bf S}_{\alpha}$
and a toroidal moment
${\bf T} \propto \sum_{\alpha} {\bf r}_{\alpha} \times {\bf S}_{\alpha}\quad$\cite{Spaldin/Fiebig/Mostovoy:2008}.

\begin{figure}[h]
  \begin{center}  
\resizebox{0.9\columnwidth}{!}{\includegraphics{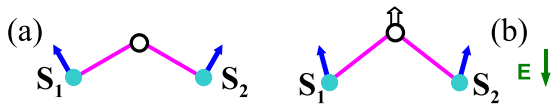}}
  \end{center}
  \caption{(a) Two magnetic cations (solid circles) connected by a ligand (open circle); (b) Upon application of electric field 
    the bond lengths and angle change resulting in a different relative alignment of spins ${\bf S_1}$ and $
{\bf S_2}$.}
  \label{fig:3atom}
\end{figure}

Next we analyze the spin-lattice coupling resulting from the dependence of the Heisenberg superexchange 
interaction between spins on their relative positions. Consider a three-atom unit consisting of two magnetic 
transition metal ions connected by a ligand such as oxygen that mediates superexchange (Fig.~\ref{fig:3atom}). 
Due to charge differences, the cations and ligand shift in opposite directions under application of an electric 
field. The exchange constant coupling the spins depends on the amplitude of the relative shifts through the
changes in the metal-oxygen distance and the metal-oxygen-metal bond angle $\theta$. According to the 
Anderson-Kanamori-Goodenough rules\cite{Anderson:Book}, the exchange is antiferromagnetic ($J>0$) for 
$\theta=180^\circ$ and ferromagnetic ($J<0$) for $\theta=90^\circ$. Experiments varying A-site cation 
size in transition metal oxides have shown that the crossover from ferromagnetic to antiferromagnetic 
coupling is continuous\cite{Subramanaian/Ramirez/Marshall:1999}. Therefore, the total magnetization of 
the unit can be modified by applying an electric field; conversely changes in spin orientation will affect 
its electric dipole moment. 

\begin{figure}[h]
\begin{center} 
\resizebox{0.9\columnwidth}{!}{\includegraphics{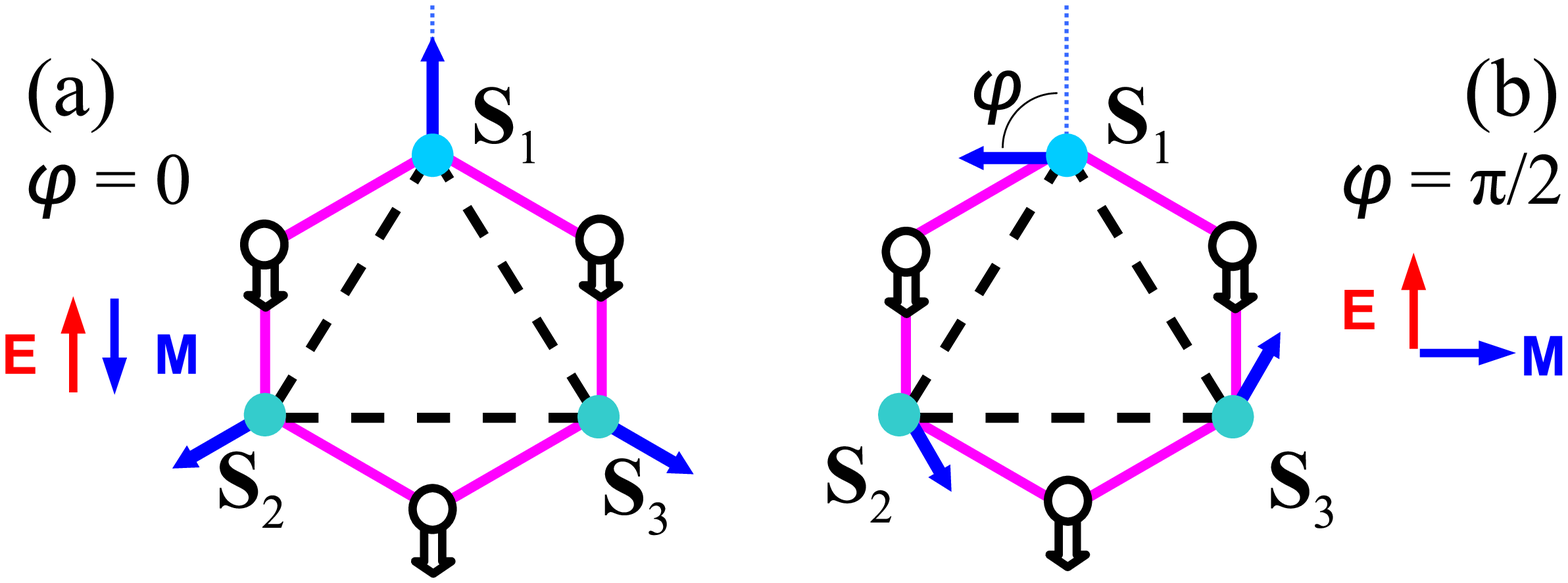}}
\end{center}
  \caption{Magnetoelectric response of a single TM-O triangular unit. For antiferromagnetic exchange coupling the 
  angle between the spins $S_1$, $S_2$ and $S_3$ (solid arrows) in the zero-field ground state is 120$^\circ$,  
  so that the net magnetization is zero. For $\varphi=0$ [panel (a)] all spins are oriented out from the center of 
  the triangle resulting in a nonzero pseudoscalar moment. Upon application of an electric field, $E$, the oxygen 
  atoms (open circles) displace (open arrows) relative to manganese (solid circles) inducing a net magnetization 
  through changes in the exchange coupling. The net magnetization $M$ is then opposite in direction to $E$ for $\varphi=0$, 
  regardless of the orientation of $E$ with respect to the spins. For $\varphi=\frac{\pi}{2}$ [panel (b)] the spin 
  triangle has a toroidal moment and the induced magnetization is perpendicular to the electric field.}
  \label{fig:triangle}
\end{figure}

Now we combine these two concepts to form a periodic array of magnetic vortices in which the magnetic
moments are coupled through superexchange, and show that the combination leads to a large magnetoelectric 
response. The macroscopic magnetoelectric response of an array of magnetic vortices is proportional to
the vortex density. Therefore we choose the smallest possible magnetic vortex as our building block: a 
triangle of antiferromagnetically coupled spins, in which the angle between spins in the lowest-energy 
state is 120$^\circ$ (Fig.~\ref{fig:triangle}a). Using transition metal (TM) ions to provide the spins,
and incorporating oxygen ligands between them (Fig.~\ref{fig:triangle}a), leads to superexchange 
spin-spin interactions. Upon application of an electric field, the shifts of the oxygen anions
relative to the positive TM ions induce changes in the Heisenberg exchange energy, changing the spin 
canting angles and resulting in a nonzero magnetization. The symmetry of the magnetoelectric response 
of the triangle is identical to that of the magnetic vortices of Fig.~\ref{fig:vortex}, with the
form of the in-plane magnetoelectric tensor constrained by its C$_{3\mathrm{v}}$ symmetry to
\begin{equation}
\alpha_{ij} = - \alpha_0
\left(
\begin{array}{cc}
\cos \varphi & \sin \varphi \\
-\sin \varphi & \cos \varphi
\end{array}
\right) \quad.
\label{eq:alpha}
\end{equation} 
Here $\varphi$ is the angle between spins and the vectors directed from the triangle center to the corresponding 
vertex. 
In particular, $\varphi=0$ (Fig.~\ref{fig:triangle}a) leads to $\alpha_{ij}=-\alpha_{0} \delta_{ij}$, so that for 
$\alpha_0 > 0$ the induced magnetization ${\bf M}$ is antiparallel to the electric field ${\bf E}$; 
for $\varphi = \pi/2$, ${\bf M}$  is perpendicular to ${\bf E}$ (Fig.~\ref{fig:triangle}b). 

\begin{figure}[h]
  \begin{center} 
\resizebox{0.8\columnwidth}{!}{\includegraphics{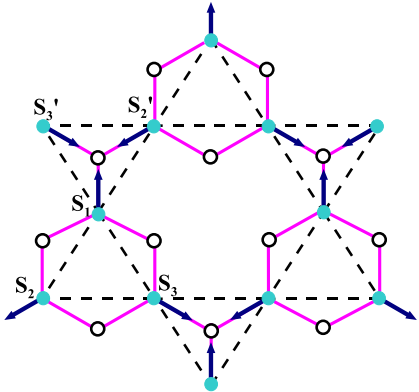}}
  \end{center}
  \caption{The structure of one MnO plane. The Mn atoms (solid circles) are 
    arranged on a Kagom\'e lattice (dashed lines) with oxygen atoms (open circles) mediating the binding and 
    superexchange.}
  \label{fig:MnOPlanes}
\end{figure}

To transform the concepts outlined above into a model material with a three-dimensional periodic structure, 
we begin with planes of Mn atoms situated on the vertices of a Kagom\'e lattice and assume that their spins form the 
$120^{\circ}$ structure with zero wave vector (see Fig. \ref{fig:MnOPlanes}), as observed e.g. in iron jarosite 
KFe$_{3}$(OH)$_{6}$(SO$_{4}$)$_{2}$\cite{Grohol_et_al:2005}. At first glance, such a spin lattice would yield no 
magnetoelectric response because spins in the vortices formed at ``up'' and ``down'' triangles are oriented in opposite 
senses ($\varphi=0$ and $\varphi=\pi$). However, when oxygen ions are positioned outside the ``up'' triangles and 
inside the ``down'' triangles, the sign of magnetoelectric coupling [$\alpha_0$ in Eq. \ref{eq:alpha}] also alternates 
and the contributions of all triangles to $\alpha_{ij}$ have the same sign. This can be understood by comparing the
magnetoelectric response of the ${\bf S}_{2}-{\bf S}_{1}$ and ${\bf S}_{1}-{\bf S}_{2}^{\prime}$ spin pairs and
noting that, for fixed bond lengths and angle, only the scalar product of spins is important in Heisenberg exchange.

\begin{figure}[h]
  \begin{center}
\resizebox{0.8\columnwidth}{!}{\includegraphics{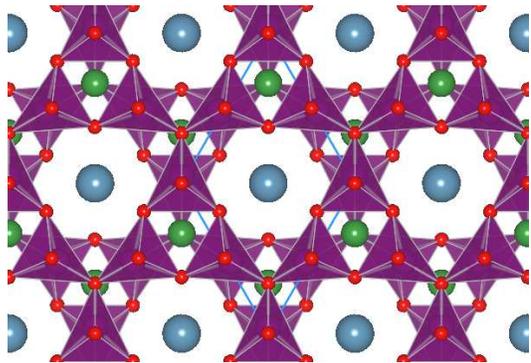}}
  \end{center}
  \caption{Two layers of KITPite in the relaxed structure (space group Pmma). The Mn ions are at the center 
    of the oxygen trigonal bipyramids (purple polyhedra). The correct charge balance is obtained through Ca (blue) 
    and Al (green) counter ions situated in voids of the MnO mesh.}
  \label{fig:3DMaterial}
\end{figure}

The two-dimensional plane shown in Fig.~\ref{fig:MnOPlanes} has a similar structure to the MnO layers of the 
experimentally realized YMnO$_3$ structure\cite{VanAken_et_al:2004}, which consists of a connected mesh of 
oxygen trigonal bipyramids with Mn atoms at their centers. Using this structure as motivation, we extend our 
two-dimensional MnO planes to a three-dimensional periodic structure and introduce counter-ions (Ca and Al) 
in the voids of the lattice so that the correct charge balance is attained.  To ensure that the sign of magnetoelectric 
response is the same for all layers, the neighboring MnO planes are rotated by 180$^\circ$ with respect to each 
other (Fig.~\ref{fig:3DMaterial}). This reverses the positioning of oxygen ions with respect to the ``up" and 
``down" spin triangles in the next layer, which compensates the reversal of the spin direction that must 
result from the antiferromagnetic interlayer coupling provided by the 180$^\circ$ connections through the apical 
oxygen atoms. Our resulting `KITPite' structure\footnote{After the Kavli Institute for Theoretical Physics, where 
this structure was first suggested by the authors.}, with chemical formula CaAlMn$_3$O$_7$, correctly breaks $I$ 
and $T$ symmetry; in addition the apical oxygen ions between the MnO layers are centers of combined $IT$ symmetry. 

Finally, to assess the strength of the magnetoelectric response of the model material introduced in 
the previous section, we turn to first principles calculations employing plane-wave density 
functional theory (DFT) as implemented in VASP\cite{Kresse/Furthmuller:1996}. We use PAW potentials 
for core-valence separation\cite{Kresse/Joubert:1999}, and include non-collinear magnetism for the valence
electrons. We approximate the exchange-correlation part of the Kohn-Sham potential using the
rotationally invariant form of the LSDA$+U$ in the fully localized 
limit\cite{Liechtenstein/Anisimov/Zaanen:1995}, with the Hubbard $U$ applied only to the Mn $d$ electrons 
($U=5.5$\,eV and $J=0.5$\,eV\cite{Yang_et_al:1999}). 
It is important to note that we deliberately do not include spin-orbit coupling in our calculations
so as to ensure that our calculated magnetoelectric response arises entirely from the superexchange 
coupling. As a result, the pseudoscalar and toroidal spin arrangements (Figs.~\ref{fig:triangle} 
(a) and (b))
are degenerate, and we are unable to determine whether $\alpha$ is diagonal or antisymmetric.

We first relax the structure in the absence of an electric field by optimizing the ionic coordinates to 
find the lowest-energy state with the constraint that the trigonal bipyramids are prevented from tilting. 
This constraint preserves the Kagom\'e structure of the MnO planes which is essential for demonstrating the 
superexchange magnetoelectricity of this model. The resulting structure is shown in Fig. \ref{fig:3DMaterial}. 
Our calculated valence electronic structure is, as expected, similar to that of YMnO$_3$\cite{VanAken_et_al:2004}:
The formal Mn charge is $3+$, with four majority $d$ electrons per Mn ($d_{xy}$, $d_{x^2-y^2}$, $d_{yz}$ and
$d_{xz}$), and an unoccupied minority channel providing a local moment of $4$\, $\mu_\mathrm{B}$/Mn. 

Subsequently, we apply an electric field and calculate the linear response of the ions, which is sufficient for
computing the linear magnetoelectric coefficient.  The force on an ion upon application 
of an external electric field is determined by the Born effective charge tensor, $Z^\star$, through 
$F_{\mu i}=Z^\star_{\mu ij}E_j$, where ${\bf F}$ is the force, ${\bf E}$ is the applied electric field, $\mu$ is 
an index denoting the ion, and $i,j$ are spatial directions. The summation convention for repeated 
indices is once again employed. All elements of the $Z^\star$ tensor are computed through derivatives 
of the bulk polarization $Z^\star_{\mu ij} = \frac{\delta P_j}{\delta R_{\mu i}}$, where ${\bf P}$ is 
calculated using the Berry-phase approach\cite{King-Smith/Vanderbilt:1993} for a small displacement 
in all degrees of freedom individually. The method is 
close to that employed by \'I\~niguez\cite{Iniguez:2008}.

In order to obtain the first-order ionic response to the field, we use the force-constant matrix 
$C_{\mu i,\nu j} = \frac{\delta F_{\mu i}}{\delta R_{\nu j}}$. Then, to linear order, the ionic 
displacements for a given force are found by inverting the force-constant matrix through 
$\delta R_{\nu j} = C_{\nu j,\mu i}^{-1} \delta F_{\mu i}$, so that the ionic response to an 
applied electric field is
$\delta R_{\nu j} = C_{\nu j,\mu i}^{-1} Z^\star_{\mu ik}E_k$.  Finally, the total magnetization 
is calculated as a function of $E_k$, yielding the linear magnetoelectric 
coupling constant.

\begin{figure}[h]
  \begin{center}  
  \resizebox{0.9\columnwidth}{!}{\includegraphics[angle=0]{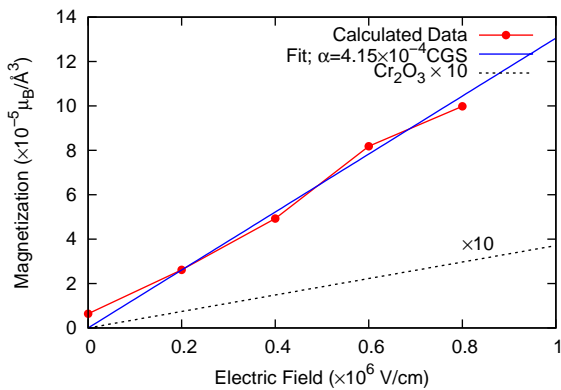}}
  \end{center}
  \caption{Calculated magnetoelectric response of the model system using density 
    functional theory and a linear fit.}
  \label{fig:MEResponse}
\end{figure}

Fig. \ref{fig:MEResponse} shows the calculated magnitude of the induced magnetization as a 
function of applied electric field. With a field of $1\times10^{6}$\,V/cm, the ionic response 
leads to an average displacement of Mn atoms of $0.007$\AA{} in the direction of the field 
and of O atoms of $0.005$\AA{} against the field. The Born effective charges, $Z^\star$, 
have an in-plane average magnitude of $+3.30 e^-$ for Mn and $-2.26 e^-$ for O. Using an equilibrium 
spin arrangement as shown in Fig. \ref{fig:MnOPlanes}, the spins $S_2$ and $S_3$ rotate by 
$-0.1^\circ$ and $0.1^\circ$ respectively, leading to a magnetoelectric coupling coefficient
of $\alpha=1.10\times10^{-5}$\,JT$^{-1}$V$^{-1}$m$^{-2}$. Transformation to regularized CGS 
units yields $\alpha_\mathrm{CGS}=4.15\times10^{-3}$. For a benchmark, we compare to the 
magnetoelectric response of Cr$_2$O$_3$ computed also within density functional theory\cite{Iniguez:2008}, 
$\alpha\left(\mathrm{Cr}_2\mathrm{O}_3\right) = 1.3\times10^{-4}$ in Gaussian units 
(in good agreement with the experimental value). Hence, our model system has a magnetoelectric 
coupling around 30 times larger than that of Cr$_2$O$_3$. Since the spin-orbit coupling was not 
considered in this work, magnetic anisotropies that determine the angle $\varphi$ were 
neglected. We therefore cannot predict whether KITPite would carry a pseudoscalar or a 
toroidal moment in the ground state. However, the strength of the magnetoelectric coupling 
resulting from Heisenberg superexchange [$\alpha_0$ in Eq.(\ref{eq:alpha})] is insensitive 
to $\varphi$. 
 
In conclusion, we have combined the concepts of magnetically induced polarization in magnetic
vortices with lattice-mediated coupling through the superexchange mechanism to demonstrate
strong magnetoelectric coupling in geometrically frustrated antiferromagnets. We showed that 
such a mechanism can be studied using modern density-functional theory approaches with non-collinear 
spin density functionals and augmented with linear-response methods, and we explicitly calculated
the magnetoelectric coupling of a model transition metal oxide. While the linear magnetoelectric
response of our model compound is larger than that of any known single phase material, we anticipate 
that many further improvements are possible: In particular materials with larger polarizability 
through increased $Z^\star$s or reduced rigidity would be promising. We hope that this study 
will stimulate the search for additional novel strongly-coupled magnetoelectric materials. 

\begin{acknowledgments}
This work was initiated during the research program on Moments and Multiplets in Mott Materials 
at the Kavli Institute for Theoretical Physics at UC Santa Barbara under the NSF grant No. PHY05-51164. 
Delaney and Spaldin were supported by the National Science Foundation under Award No. DMR-0605852. 
Calculations were performed at the UCSB California 
Nanosystems Institute (CNSI) with facilities provided by NSF Award No. CHE-0321368 and 
Hewlett-Packard, at the San Diego Supercomputer Center, and at the National Center for Supercomputer Applications.
\end{acknowledgments}

\bibliography{kitpite}

\begin{thebibliography}{21}
\expandafter\ifx\csname natexlab\endcsname\relax\def\natexlab#1{#1}\fi
\expandafter\ifx\csname bibnamefont\endcsname\relax
  \def\bibnamefont#1{#1}\fi
\expandafter\ifx\csname bibfnamefont\endcsname\relax
  \def\bibfnamefont#1{#1}\fi
\expandafter\ifx\csname citenamefont\endcsname\relax
  \def\citenamefont#1{#1}\fi
\expandafter\ifx\csname url\endcsname\relax
  \def\url#1{\texttt{#1}}\fi
\expandafter\ifx\csname urlprefix\endcsname\relax\def\urlprefix{URL }\fi
\providecommand{\bibinfo}[2]{#2}
\providecommand{\eprint}[2][]{\url{#2}}

\bibitem[{\citenamefont{Landau and Lifshitz}(1960)}]{Landau/Lifshitz:Book}
\bibinfo{author}{\bibfnamefont{L.~D.} \bibnamefont{Landau}} \bibnamefont{and}
  \bibinfo{author}{\bibfnamefont{E.~M.} \bibnamefont{Lifshitz}},
  \emph{\bibinfo{title}{Electrodynamics of Continuous Media}}
  (\bibinfo{publisher}{Addison-Wesley, Reading, MA, USA},
  \bibinfo{year}{1960}).

\bibitem[{\citenamefont{Fiebig}(2005)}]{Fiebig:2005}
\bibinfo{author}{\bibfnamefont{M.}~\bibnamefont{Fiebig}},
  \bibinfo{journal}{J.~Phys. D: Appl. Phys.} \textbf{\bibinfo{volume}{38}},
  \bibinfo{pages}{R123} (\bibinfo{year}{2005}).

\bibitem[{\citenamefont{Newnham et~al.}(1978)\citenamefont{Newnham, Kramer,
  Schulze, and Cross}}]{Newnham/Kramer/Schulze/Cross_1978}
\bibinfo{author}{\bibfnamefont{R.~E.} \bibnamefont{Newnham}},
  \bibinfo{author}{\bibfnamefont{J.~J.} \bibnamefont{Kramer}},
  \bibinfo{author}{\bibfnamefont{W.~A.} \bibnamefont{Schulze}},
  \bibnamefont{and} \bibinfo{author}{\bibfnamefont{L.~E.} \bibnamefont{Cross}},
  \bibinfo{journal}{J.~Appl. Phys.} \textbf{\bibinfo{volume}{49}},
  \bibinfo{pages}{12} (\bibinfo{year}{1978}).

\bibitem[{\citenamefont{Kimura et~al.}(2003)\citenamefont{Kimura, Goto,
  Shintani, Ishizaka, Arima, and Tokura}}]{Kimura_et_al:2003}
\bibinfo{author}{\bibfnamefont{T.}~\bibnamefont{Kimura}},
  \bibinfo{author}{\bibfnamefont{T.}~\bibnamefont{Goto}},
  \bibinfo{author}{\bibfnamefont{H.}~\bibnamefont{Shintani}},
  \bibinfo{author}{\bibfnamefont{K.}~\bibnamefont{Ishizaka}},
  \bibinfo{author}{\bibfnamefont{T.}~\bibnamefont{Arima}}, \bibnamefont{and}
  \bibinfo{author}{\bibfnamefont{Y.}~\bibnamefont{Tokura}},
  \bibinfo{journal}{Nature} \textbf{\bibinfo{volume}{426}}, \bibinfo{pages}{55}
  (\bibinfo{year}{2003}).

\bibitem[{\citenamefont{Goto et~al.}(2004)\citenamefont{Goto, Kimura, Lawes,
  Ramirez, and Tokura}}]{Goto_et_al:2004}
\bibinfo{author}{\bibfnamefont{T.}~\bibnamefont{Goto}},
  \bibinfo{author}{\bibfnamefont{T.}~\bibnamefont{Kimura}},
  \bibinfo{author}{\bibfnamefont{G.}~\bibnamefont{Lawes}},
  \bibinfo{author}{\bibfnamefont{A.~P.} \bibnamefont{Ramirez}},
  \bibnamefont{and} \bibinfo{author}{\bibfnamefont{Y.}~\bibnamefont{Tokura}},
  \bibinfo{journal}{Phys. Rev. Lett.} \textbf{\bibinfo{volume}{92}},
  \bibinfo{pages}{257201} (\bibinfo{year}{2004}).

\bibitem[{\citenamefont{Katsura et~al.}(2005)\citenamefont{Katsura, Nagaosa,
  and Balatsky}}]{Katsura/Nagaosa/Balatsky:2005}
\bibinfo{author}{\bibfnamefont{H.}~\bibnamefont{Katsura}},
  \bibinfo{author}{\bibfnamefont{N.}~\bibnamefont{Nagaosa}}, \bibnamefont{and}
  \bibinfo{author}{\bibfnamefont{A.~V.} \bibnamefont{Balatsky}},
  \bibinfo{journal}{Phys. Rev. Lett.} \textbf{\bibinfo{volume}{95}},
  \bibinfo{pages}{057205} (\bibinfo{year}{2005}).

\bibitem[{\citenamefont{Sergienko and Dagotto}(2006)}]{Sergienko/Dagotto:2006}
\bibinfo{author}{\bibfnamefont{I.~A.} \bibnamefont{Sergienko}}
  \bibnamefont{and} \bibinfo{author}{\bibfnamefont{E.}~\bibnamefont{Dagotto}},
  \bibinfo{journal}{Phys. Rev. B} \textbf{\bibinfo{volume}{73}},
  \bibinfo{pages}{094434} (\bibinfo{year}{2006}).

\bibitem[{\citenamefont{Sergienko et~al.}(2006)\citenamefont{Sergienko, Sen,
  and Dagotto}}]{Sergienko/Sen/Dagotto:2006}
\bibinfo{author}{\bibfnamefont{I.~A.} \bibnamefont{Sergienko}},
  \bibinfo{author}{\bibfnamefont{C.}~\bibnamefont{Sen}}, \bibnamefont{and}
  \bibinfo{author}{\bibfnamefont{E.}~\bibnamefont{Dagotto}},
  \bibinfo{journal}{Phys. Rev. Lett.} \textbf{\bibinfo{volume}{97}},
  \bibinfo{pages}{227204} (\bibinfo{year}{2006}).

\bibitem[{\citenamefont{Picozzi et~al.}(2007)\citenamefont{Picozzi, Yamaguchi,
  Sanyal, Sergienko, and Dagotto}}]{Picozzi_et_al:2007}
\bibinfo{author}{\bibfnamefont{S.}~\bibnamefont{Picozzi}},
  \bibinfo{author}{\bibfnamefont{K.}~\bibnamefont{Yamaguchi}},
  \bibinfo{author}{\bibfnamefont{B.}~\bibnamefont{Sanyal}},
  \bibinfo{author}{\bibfnamefont{I.~A.} \bibnamefont{Sergienko}},
  \bibnamefont{and} \bibinfo{author}{\bibfnamefont{E.}~\bibnamefont{Dagotto}},
  \bibinfo{journal}{Phys. Rev. Lett.} \textbf{\bibinfo{volume}{99}},
  \bibinfo{pages}{227201} (\bibinfo{year}{2007}).

\bibitem[{\citenamefont{Mostovoy}(2006)}]{Mostovoy:2006}
\bibinfo{author}{\bibfnamefont{M.}~\bibnamefont{Mostovoy}},
  \bibinfo{journal}{Phys. Rev. Lett.} \textbf{\bibinfo{volume}{96}},
  \bibinfo{pages}{067601} (\bibinfo{year}{2006}).

\bibitem[{\citenamefont{Spaldin et~al.}()\citenamefont{Spaldin, Fiebig, and
  Mostovoy}}]{Spaldin/Fiebig/Mostovoy:2008}
\bibinfo{author}{\bibfnamefont{N.~A.} \bibnamefont{Spaldin}},
  \bibinfo{author}{\bibfnamefont{M.}~\bibnamefont{Fiebig}}, \bibnamefont{and}
  \bibinfo{author}{\bibfnamefont{M.}~\bibnamefont{Mostovoy}},
  \bibinfo{journal}{J.~Phys.: Condens. Matter} \textbf{\bibinfo{volume}{in
  press}} (????).

\bibitem[{\citenamefont{Anderson}(1963)}]{Anderson:Book}
\bibinfo{author}{\bibfnamefont{P.~W.} \bibnamefont{Anderson}},
  \emph{\bibinfo{title}{Magnetism}}, vol.~\bibinfo{volume}{1}
  (\bibinfo{publisher}{Academic Press, New York}, \bibinfo{year}{1963}).

\bibitem[{\citenamefont{Subramanian et~al.}(1999)\citenamefont{Subramanian,
  Ramirez, and Marshall}}]{Subramanaian/Ramirez/Marshall:1999}
\bibinfo{author}{\bibfnamefont{M.~A.} \bibnamefont{Subramanian}},
  \bibinfo{author}{\bibfnamefont{A.~P.} \bibnamefont{Ramirez}},
  \bibnamefont{and} \bibinfo{author}{\bibfnamefont{W.~J.}
  \bibnamefont{Marshall}}, \bibinfo{journal}{Phys. Rev. Lett.}
  \textbf{\bibinfo{volume}{82}}, \bibinfo{pages}{1558} (\bibinfo{year}{1999}).

\bibitem[{\citenamefont{Grohol et~al.}(2005)\citenamefont{Grohol, Matan, Cho,
  Lee, Lynn, Nocera, and Lee}}]{Grohol_et_al:2005}
\bibinfo{author}{\bibfnamefont{D.}~\bibnamefont{Grohol}},
  \bibinfo{author}{\bibfnamefont{K.}~\bibnamefont{Matan}},
  \bibinfo{author}{\bibfnamefont{J.-H.} \bibnamefont{Cho}},
  \bibinfo{author}{\bibfnamefont{S.-H.} \bibnamefont{Lee}},
  \bibinfo{author}{\bibfnamefont{J.~W.} \bibnamefont{Lynn}},
  \bibinfo{author}{\bibfnamefont{D.~G.} \bibnamefont{Nocera}},
  \bibnamefont{and} \bibinfo{author}{\bibfnamefont{Y.~S.} \bibnamefont{Lee}},
  \bibinfo{journal}{Nature Materials} \textbf{\bibinfo{volume}{4}},
  \bibinfo{pages}{323} (\bibinfo{year}{2005}).

\bibitem[{\citenamefont{Van~Aken et~al.}(2004)\citenamefont{Van~Aken, Palstra,
  Filippetti, and Spaldin}}]{VanAken_et_al:2004}
\bibinfo{author}{\bibfnamefont{B.~B.} \bibnamefont{Van~Aken}},
  \bibinfo{author}{\bibfnamefont{T.~T.~M.} \bibnamefont{Palstra}},
  \bibinfo{author}{\bibfnamefont{A.}~\bibnamefont{Filippetti}},
  \bibnamefont{and} \bibinfo{author}{\bibfnamefont{N.~A.}
  \bibnamefont{Spaldin}}, \bibinfo{journal}{Nature Materials}
  \textbf{\bibinfo{volume}{3}}, \bibinfo{pages}{164} (\bibinfo{year}{2004}).

\bibitem[{\citenamefont{Kresse and
  Furthm{\"u}ller}(1996)}]{Kresse/Furthmuller:1996}
\bibinfo{author}{\bibfnamefont{G.}~\bibnamefont{Kresse}} \bibnamefont{and}
  \bibinfo{author}{\bibfnamefont{J.}~\bibnamefont{Furthm{\"u}ller}},
  \bibinfo{journal}{Phys. Rev. B} \textbf{\bibinfo{volume}{54}},
  \bibinfo{pages}{11169} (\bibinfo{year}{1996}).

\bibitem[{\citenamefont{Kresse and Joubert}(1999)}]{Kresse/Joubert:1999}
\bibinfo{author}{\bibfnamefont{G.}~\bibnamefont{Kresse}} \bibnamefont{and}
  \bibinfo{author}{\bibfnamefont{D.}~\bibnamefont{Joubert}},
  \bibinfo{journal}{Phys. Rev. B} \textbf{\bibinfo{volume}{59}},
  \bibinfo{pages}{1758} (\bibinfo{year}{1999}).

\bibitem[{\citenamefont{Liechtenstein et~al.}(1995)\citenamefont{Liechtenstein,
  Anisimov, and Zaanen}}]{Liechtenstein/Anisimov/Zaanen:1995}
\bibinfo{author}{\bibfnamefont{A.~I.} \bibnamefont{Liechtenstein}},
  \bibinfo{author}{\bibfnamefont{V.~I.} \bibnamefont{Anisimov}},
  \bibnamefont{and} \bibinfo{author}{\bibfnamefont{J.}~\bibnamefont{Zaanen}},
  \bibinfo{journal}{Phys. Rev. B} \textbf{\bibinfo{volume}{52}},
  \bibinfo{pages}{R5467} (\bibinfo{year}{1995}).

\bibitem[{\citenamefont{Yang et~al.}(1999)\citenamefont{Yang, Huang, Ye, and
  Xie}}]{Yang_et_al:1999}
\bibinfo{author}{\bibfnamefont{Z.}~\bibnamefont{Yang}},
  \bibinfo{author}{\bibfnamefont{Z.}~\bibnamefont{Huang}},
  \bibinfo{author}{\bibfnamefont{L.}~\bibnamefont{Ye}}, \bibnamefont{and}
  \bibinfo{author}{\bibfnamefont{X.}~\bibnamefont{Xie}},
  \bibinfo{journal}{Phys. Rev. B} \textbf{\bibinfo{volume}{60}},
  \bibinfo{pages}{15674} (\bibinfo{year}{1999}).

\bibitem[{\citenamefont{King-Smith and
  Vanderbilt}(1993)}]{King-Smith/Vanderbilt:1993}
\bibinfo{author}{\bibfnamefont{R.~D.} \bibnamefont{King-Smith}}
  \bibnamefont{and}
  \bibinfo{author}{\bibfnamefont{D.}~\bibnamefont{Vanderbilt}},
  \bibinfo{journal}{Phys. Rev. B} \textbf{\bibinfo{volume}{47}},
  \bibinfo{pages}{1651} (\bibinfo{year}{1993}).

\bibitem[{\citenamefont{{\'I}{\~n}iguez}(2008)}]{Iniguez:2008}
\bibinfo{author}{\bibfnamefont{J.}~\bibnamefont{{\'I}{\~n}iguez}},
  \bibinfo{journal}{Phys. Rev. Lett.} \textbf{\bibinfo{volume}{101}},
  \bibinfo{pages}{117201} (\bibinfo{year}{2008}).

\end{thebibliography}

\end{document}